\documentclass[11pt]{article}
\usepackage{geometry}
\usepackage{color}

\usepackage{array}
\usepackage{amssymb}
\usepackage{graphicx}
\usepackage{authblk}

\makeatletter
\newcommand{\lyxdot}{.}

\title{
\begin{flushright}
\small
RIKEN-QHP-179, 
UTHEP-671, 
YITP-15-13
\end{flushright}
$\Omega\Omega$ interaction from  2+1 flavor lattice QCD}

\author[1]{Masanori~Yamada}
\author[2]{Kenji~Sasaki}
\author[2,3]{Sinya~Aoki}
\author[4]{Takumi~Doi}
\author[4,5]{Tetsuo~Hatsuda}
\author[4]{Yoichi~Ikeda}
\author[6]{Takashi~Inoue}
\author[7]{Noriyoshi~Ishii}
\author[7]{Keiko~Murano}
\author[1,2]{Hidekatsu~Nemura}
\author[]{(HAL QCD Collaboration)}

\affil[1]{Graduate School of Pure and Applied Sciences, University of Tsukuba, Tsukuba, 305-8571, Japan}
\affil[2]{Center for Computational Sciences, University of Tsukuba, Tsukuba, 305-8577, Japan}
\affil[3]{Yukawa Institute for Theoretical Physics, Kyoto University, Kyoto, 606-8502, Japan}
\affil[4]{Theoretical Research Division, Nishina Center, RIKEN, Wako, 351-0198, Japan}
\affil[5]{Kavli IPMU (WPI), The University of Tokyo, Kashiwa, Chiba 277-8583, Japan}
\affil[6]{Nihon University,  College of Bioresource Sciences, Fujisawa, 252-0880, Japan}
\affil[7]{Research Center for Nuclear Physics (RCNP), Osaka University, Ibaraki, 567-0047, Japan}

\begin{document}
\maketitle
\makeatother

\begin{abstract}
%---200 words limit for PTEP Letter ---
We investigate the interaction between $\Omega$ baryons in the $^1S_0$ 
channel from 2+1 flavor lattice QCD simulations. On the basis of the 
HAL QCD method, the $\Omega\Omega$ potential is extracted 
from the Nambu-Bethe-Salpeter wave function  calculated on 
the lattice by using the PACS-CS gauge configurations
 with the lattice spacing $a\simeq 0.09$ fm, 
 the lattice volume $L\simeq 2.9$ fm and the quark masses 
 corresponding to $m_\pi \simeq 700$ MeV and 
$m_\Omega \simeq 1970$ MeV.  The $\Omega\Omega$ potential has a repulsive core at short 
distance and an attractive well at intermediate distance. Accordingly, the phase shift 
 obtained from the potential shows moderate attraction at low energies.
 Our data indicate that the  $\Omega\Omega$ system with the present quark masses 
  may appear close to the  unitary limit where the scattering length diverges.
\end{abstract}

%\subjectindex{D32, D34, B64}

{\bf\em Introduction}\ \ 
Strange dibaryons have been attracting considerable interests both theoretically and experimentally
 in hadron physics. In particular, the $H$-dibaryon with (strangeness)=$-$2 \cite{Jaffe1977} 
 and the $N\Omega$ dibaryon with (strangeness)=$-$3 \cite{Goldman:1987ma} are considered to be 
  the promising dibaryon states  due to the absence of 
 Pauli repulsions among valence quarks at short distance (see the reviews, \cite{Oka:1988yq,Gal:2010eg}).
  In recent years, the numerical and theoretical progresses in lattice gauge theories made
  it  possible to attack such a problem directly from quantum chromodynamics (QCD)
  (See e.g.\cite{Inoue:2011ai,Beane:2011iw,Haidenbauer:2014rna} and references therein.) 
  
 The purpose of this letter is to extend our previous works on the (strangeness)=$-$2  systems such as $H$-dibaryon~\cite{Inoue:2011ai} and $N\Xi$~\cite{Nemura:2008sp} and the (strangeness)=$-$3  system as $N\Omega$ \cite{Etminan:2014tya} 
 to the (strangeness)=$-$6  $\Omega\Omega$ system in 2+1 flavor lattice QCD.
 In our approach (the HAL QCD method), the baryon-baryon potential is extracted from the 
 Nambu-Bethe-Salpeter (NBS) wave function calculated on the lattice:  Such a 
 potential deduced in lattice QCD is guaranteed to reproduce physical observables 
 (e.g. the scattering phase shift) by construction \cite{Ishii:2006ec}.  
 The HAL QCD method has several advantages 
  over the conventional finite volume method \cite{luscher:1991}: 
   (i) The finite volume effect is highly suppressed due to the short range character of 
   baryon potentials, (ii) the ground state saturation of the two-particle system 
   is not required for extracting the potential, since the same potential distates
   all the scattering states on the lattice, and (iii) physics behind the baryon-baryon interaction
    can be easily grasped from the spatial structure of the potential. 
    Further details  on these points are discussed in  \cite{Ishii:2012plb,Aoki:2012tk}. 
    
 There exit several investigations so far on the $\Omega\Omega$ interaction
 using  the phenomenological quark models:
 Some studies show strong attraction which may cause a $\Omega\Omega$ bound state
\cite{zhang:1997,zhang:2000}, while other studies show weak repulsion \cite{wang:1992,wang:1997}.
A recent lattice QCD analysis of the $\Omega\Omega$ scattering length
\cite{Buchoff:2012prd} by using the standard  finite volume method \cite{luscher:1991} shows 
weak repulsion in the $S$-wave scattering with the scattering  length $a_{\Omega\Omega} = - 0.16 \pm 0.22$fm: 
This indicates that the $\Omega\Omega$ system is unlikely to have a strongly bound dibaryon, although 
the large error  prevents us to make a firm conclusion about details of the interaction.
 
 \
 
\noindent
{\bf\em The HAL QCD potential} \ \ 
Let us first recapitulate the essential part of  the HAL QCD method to be used for extracting the  $\Omega\Omega$ potential.
The basic quantity is the equal-time NBS wave function with  the Euclidean time $t$,
\begin{equation}
\psi_{\alpha k,\beta l}^{W}(\vec{r})e^{-W t}\equiv\left\langle 0\right\vert \Omega_{\alpha,k}(t,\vec{r})\Omega_{\beta,l}(t,\vec{0}) \left\vert \Omega\Omega, W \right\rangle ,
\end{equation}
where $\left\vert\Omega\Omega, W  \right\rangle$ is
  an exact (strangeness)$=-6$ state with zero total momentum.
The  total energy of the system
is given by $W=2\sqrt{m_{\Omega}^{2}+\vec p^{2}}$ with the $\Omega$ baryon mass $m_\Omega$ and the relative momentum $\vec p$.
Local interpolating operators for the $\Omega$ baryon, $\Omega(x)$ and $\overline{\Omega}(x)$, are taken to be 
\begin{eqnarray}
\Omega_{\alpha,k}(x) &\equiv& \varepsilon^{abc}s_{a}^{T}(x)C\gamma_{k}s_{b}(x)s_{c,\alpha}(x), \nonumber \\
\overline{\Omega}_{\alpha,k}(x) &\equiv& \Omega_{\alpha,k}^{\dagger}\gamma^{0} =-\varepsilon^{abc}\overline{s}_{a,\alpha}(x)\overline{s}_{b}(x)\gamma_{k_{1}}C\overline{s}^{T}_{c}(x),
\end{eqnarray}
where $a$, $b$ and $c$ are color indices, $\varepsilon^{abc}$ is the totally anti-symmetric tensor,
$\gamma_{k}$ represents the Dirac matrix, and $\alpha$ is the spinor  index. The
charge conjugation matrix  is taken as $C\equiv\gamma_{4}\gamma_{2}=-C^{-1}=-C^{T}=-C^{\dagger}$ in the Euclidean space. 

An important property of the NBS wave function is its asymptotic behavior at large distance, denoted simply as 
\begin{equation}
\psi^{W} (\vec r) \sim \sum_{L,M} e^{i\delta_L(p)}\frac{\sin(pr-\frac{L\pi}{2}+\delta_L(p))}{pr} C_{L,M} Y_{LM}(\vec\Omega),
\label{eq:NBS-asymp}
\end{equation}
where $p=\vert\vec p\vert$, $r=\vert\vec r\vert$, $\vec\Omega$ is the solid angle of $\vec r$, $Y_{LM}$ is the spherical harmonic function,  and $L$ is the orbital angular momentum.
Although Eq.(\ref{eq:NBS-asymp}) looks like a quantum mechanical formula,  
it can be derived from the  unitarity of the $S$-matrix in quantum field theory with 
 $\delta_L(p)$ being the scattering phase shift for given quantum numbers in QCD \cite{Ishii:2006ec}.

Our next task is to define the potential from which this scattering phase shift can be calculated.
In the HAL QCD method, such a  potential is defined through  the Schr\"{o}dinger type equation,
\begin{equation}
(E-H_{0})\psi^{W}(\vec r)=\int d^{3}r'U(\vec r,\vec r')\psi^{W}(\vec r'),\label{eq:sch}
\end{equation}
where $H_{0}\equiv-\frac{1}{2\mu_{\Omega}}\nabla^{2}$ is the
free Hamiltonian, $\mu_{\Omega}\equiv m_{\Omega}/2$ is the reduced mass, and 
$E\equiv\frac{1}{2\mu_{\Omega}}p^{2}$ is the kinetic energy. 
Here $U(\vec r,\vec r')$ is the non-local but energy independent  potential, which can be
 expanded in terms of the non-locality (the velocity  or derivative  expansion) \cite{Okubo:1958}.
The convergence of the velocity expansion at low energies has been investigated previously
for the nucleon-nucleon scattering \cite{murano:2011ptp}
and the pion-pion  scattering \cite{Kurth:2013tua}.
 Since we consider the low-energy $S$-wave scattering much below the meson production threshold in this letter,
 we take only the leading-order local potential, $V(\vec{r})$,  in the expansion,

$U(\vec{r},\vec{r}')=V(\vec{r})\delta(\vec{r}-\vec{r}')+\mathcal{O}(\vec{\nabla}).$

Note that  $V(\vec r)$ is an  effective central potential, which contains 
 not only the genuine central part but also  the tensor part implicitly \cite{Ishii:2006ec}. 

The NBS wave function can be extracted from the asymptotic temporal behavior of the four-point (4-pt) function,
\begin{eqnarray}
F(\vec r, t-t_0) &=& \langle 0 \vert \Omega(t,\vec r)  \Omega(t,\vec 0)  {\mathcal J}(t_0) \vert 0 \rangle 
= \sum_n  \langle 0 \vert \Omega(t,\vec r)  \Omega(t,\vec 0)  \vert \Omega\Omega, W_n\rangle \langle \Omega\Omega, W_n \vert  {\mathcal J}(t_0) \vert 0 \rangle +\cdots
 \nonumber \\
&=& \sum_n  a_n \psi^{W_n}(\vec r) e^{-W_n(t-t_0)} +\cdots \simeq a_0 \psi^{W_0}(\vec r) e^{-W_0(t-t_0)} ,
\quad (t-t_0\rightarrow\infty), 
\label{eq:4pt} 
\end{eqnarray}
where ${\mathcal J}(t_0)$ is the wall source operator  which creates the $\Omega\Omega$ state at $t_0$,
$a_n$ is the matrix element defined by  $\langle \Omega\Omega, W_n \vert  {\mathcal J}(0) \vert 0 \rangle$, 
$W_n$ are discrete QCD eigen-energies  in the finite volume below inelastic threshold, and $W_0$  is the lowest eigen-energy. The ellipses in the above equation represent inelastic contributions in the $\Omega\Omega$ system, which are suppressed for large $t-t_0$.

Eq.~(\ref{eq:4pt}) shows that the NBS wave function for the ground state can be extracted from the 4-pt function at large $t$, 
as long as other contributions from $W_{n\ge1} > W_0$ can be neglected. 
In practice, however, the increasing statistical errors of the 4-pt function at large $t$ make it difficult to achieve the
ground state saturation with reasonable accuracy \cite{Lepage:1990}. Moreover,   
as the volume increases, larger and larger $t$ becomes necessary to extract $\psi^{W_0}(\vec r)$.
These problems can be simultaneously avoided by the time-dependent HAL QCD method introduced in \cite{Ishii:2012plb}, where
we start with the so-called  $R$-correlator,
\begin{equation}
R(\vec r, t-t_{0})\equiv \frac{F(\vec r,t-t_0)}{e^{-2m_\Omega(t-t_{0})}}=\sum_{n}a_{n}\psi^{W_n}(\vec r)e^{-\Delta W_{n}(t-t_{0})} +\cdots ,
\label{eq:R-col}
\end{equation}
with $\Delta W_n = W_n - 2 m_\Omega$. 
Since $\Delta W_{n}=\displaystyle \frac{\vec p_n^2}{m_\Omega}-\frac{\Delta W_{n}^{2}}{4m_\Omega}$,
we have
\begin{eqnarray}
-\frac{\partial}{\partial t}R(\vec r,  t)&\simeq& \sum_{n}\Delta W_{n}a_{n}\psi^{W_n}(\vec r)e^{-\Delta W_{n} t}
=\sum_{n}(\frac{\vec p_n^2}{m_\Omega}-\frac{1}{4m_\Omega}\frac{\partial^{2}}{\partial t^{2}})a_{n}\phi^{W_n}(\vec r)e^{-\Delta W_{n}t} \nonumber \\
&=& (-\frac{1}{m_\Omega}\nabla^{2}-\frac{1}{4m_\Omega}\frac{\partial ^{2}}{\partial t^{2}})R(\vec r,t)+\int d^3r'U(\vec r,\vec r')R(\vec r', t), 
\end{eqnarray}
where we  have replaced ${\vec p_{n}^{2}}/{m_\Omega}$ by
$U-\nabla^{2}/{m_\Omega}$ using Eq.(\ref{eq:sch}).
We then arrive at the time-dependent
equation,
\begin{equation}
(\frac{1}{m_\Omega}\triangledown^{2}-\frac{\partial}{\partial t}+\frac{1}{4m_\Omega}\frac{\partial^{2}}{\partial t^{2}})R(\vec r, t)
\simeq \int d^3r' U(\vec r,\vec r')R(\vec r', t),  \label{eq:tdep}  
\end{equation}
where ''$\simeq$" implies that we have neglected inelastic contributions by taking sufficiently large $t-t_0$. 
Eq.(\ref{eq:tdep})  gives  $U(\vec r,\vec r')$ directly from $F(\vec r, t)$.
This method no more requires the ground state saturation, so that
the moderately large values of $t-t_0$ which suppress inelastic contributions
are enough for a reliable extraction of the potential. 
Then, in the leading order of the velocity expansion, we obtain 
\begin{eqnarray}
V(\vec r) &=&\frac{ \left(\frac{1}{m_\Omega}\triangledown^{2}-\frac{\partial}{\partial t}+\frac{1}{4m_\Omega}\frac{\partial^{2}}{\partial t^{2}}\right)R(\vec r, t)}{ R(\vec r, t)} . \label{eq:Potential}
\end{eqnarray}

\

\noindent
{\bf \em  Interpolating operators for $\Omega\Omega$ system} \ \ 
The present system can be characterized by the total spin $(S)$, 
the orbital angular momentum $(L)$, the total angular momentum $(J)$, and the parity $P$. 
The asymptotic $\Omega\Omega$ state with given $L$ and $S$ has a factor $(-1)^{S+L+1}$ under the
exchange of two $\Omega$'s, so that $S+L$ must be even due to the Fermi statistics of $\Omega$ baryons.
In Table~\ref{tab:state}, we show low-$J$ channels $^{2S+1}L_J$ which appear for given  
conserved quantum numbers $J$ and $P$.
In this letter, we employ the wall source, $L=0$ with $S=0$ at $t=t_0$, which creates  the $J^{P}=0^{+}$  state, so that only the upper left corner of this table is considered. 
Then, both $^1S_0$ and $^5D_0$ channels appear after the QCD interactions at $t > t_0$.
 As mentioned before, we determine only the effective central potential
 from the  $^1S_0$ channel at $t > t_0$, where
 effects of the $^5D_0$ state are implicitly included.

{\footnotesize{}}
\begin{table}[t]
{\footnotesize{}}
\begin{center}
\begin{tabular}{|c|>{\centering}p{6cm}|>{\centering}p{6cm}|}
\hline 
 & {\footnotesize{}$P=+$} & {\footnotesize{}$P=-$}\tabularnewline
\hline 
{\footnotesize{}$J=0$} & {\footnotesize{}$^{1}S_{0}$, $^{5}D_{0}$} & {\footnotesize{}$^{3}P_{0}$, $^{7}F_{0}$ }\tabularnewline
\hline 
{\footnotesize{}$J=1$} & {\footnotesize{}$^{5}D_{1}$} & {\footnotesize{}$^{3}P_{1}$, $^{7}F_{1}$}\tabularnewline
\hline 
{\footnotesize{}$J=2$} & {\footnotesize{}$^{5}S_{2}$, $^{1}D_{2}$, $^{5}D_{2}$, $^{5}G_{2}$} & {\footnotesize{}$^{3}P_{2}$, $^{7}P_{2}$, $^{3}F_{2}$, $^{7}F_{2}$
, $^{7}H_{2}$}\tabularnewline
\hline 
{\footnotesize{}$J=3$} & {\footnotesize{}$^{5}D_{3}$, $^{5}G_{3}$} & {\footnotesize{}$^{7}P_{3}$, $^{3}F_{3}$, $^{7}F_{3}$ , $^{7}H_{3}$}\tabularnewline
\hline 
{\footnotesize{}$J=4$} & {\footnotesize{}$^{5}D_{4}$, $^{1}G_{4}$, $^{5}G_{4}$, $^{5}I_{4}$} & {\footnotesize{}$^{7}P_{4}$, $^{3}F_{4}$, $^{7}F_{4}$, $^{3}H_{4}$,
$^{7}H_{4}$, $^{7}J_{4}$}\tabularnewline
\hline 
\end{tabular}{\footnotesize \par}
\end{center}
\caption{A relation between conserved quantum numbers ($J$ and $P$) and quantum numbers in the asymptotic $\Omega\Omega$ channel.}
\label{tab:state}
\end{table}{\footnotesize \par}

The interpolating operators for $\Omega$ with $S=3/2$ and $S_z=\pm 3/2, \pm 1/2$ read
\begin{eqnarray}
\Omega_{\frac{3}{2},\frac{3}{2}}&=&-(\psi\Gamma_{+}\psi)\psi_{\frac{1}{2}}, \\
\Omega_{\frac{3}{2},\frac{1}{2}}&=&\frac{1}{\sqrt{3}}[\sqrt{2}(\psi\Gamma_{Z}\psi)\psi_{\frac{1}{2}}+(\psi\Gamma_{+}\psi)\psi_{-\frac{1}{2}}], \\
\Omega_{\frac{3}{2},-\frac{1}{2}}&=&\frac{1}{\sqrt{3}}[\sqrt{2}(\psi\Gamma_{Z}\psi)\psi_{-\frac{1}{2}}+(\psi\Gamma_{-}\psi)\psi_{\frac{1}{2}}], \\
\Omega_{\frac{3}{2},-\frac{3}{2}}&=&(\psi\Gamma_{-}\psi)\psi_{-\frac{1}{2}},
\end{eqnarray}
where $\Gamma_{\pm}\equiv\frac{1}{2}(C\gamma^{2}\pm iC\gamma^{1})$ and 
$\Gamma_{Z}\equiv\frac{-i}{\sqrt{2}}C\gamma^{3}$.
We take only upper two components in the Dirac representation for the quark operators, so that 
 the $\Omega$ operator does not have the $S=1/2$ component.
Combining these operators, the interpolating operators for the $\Omega\Omega$ system with the total spin $S=3,2,1,0$ with $S_z=0$ are given by 
\begin{eqnarray}
\label{eq:OO_30}
(\Omega\Omega)_{3,0}&=&\frac{1}{\sqrt{20}}(\Omega_{\frac{3}{2},\frac{3}{2}}\Omega_{\frac{3}{2},-\frac{3}{2}}+3\Omega_{\frac{3}{2},\frac{1}{2}}\Omega_{\frac{3}{2},-\frac{1}{2}}+3\Omega_{\frac{3}{2},-\frac{1}{2}}\Omega_{\frac{3}{2}.\frac{1}{2}}+\Omega_{\frac{3}{2},-\frac{3}{2}}\Omega_{\frac{3}{2},\frac{3}{2}}),\\
\label{eq:OO_20}
(\Omega\Omega)_{2,0}&=&\frac{1}{2}(\Omega_{\frac{3}{2},\frac{3}{2}}\Omega_{\frac{3}{2},-\frac{3}{2}}+\Omega_{\frac{3}{2},\frac{1}{2}}\Omega_{\frac{3}{2},-\frac{1}{2}}-\Omega_{\frac{3}{2},-\frac{1}{2}}\Omega_{\frac{3}{2}.\frac{1}{2}}-\Omega_{\frac{3}{2},-\frac{3}{2}}\Omega_{\frac{3}{2},\frac{3}{2}}),\\
\label{eq:OO_10}
(\Omega\Omega)_{1,0}&=&\frac{1}{\sqrt{20}}(3\Omega_{\frac{3}{2},\frac{3}{2}}\Omega_{\frac{3}{2},-\frac{3}{2}}-\Omega_{\frac{3}{2},\frac{1}{2}}\Omega_{\frac{3}{2},-\frac{1}{2}}-\Omega_{\frac{3}{2},-\frac{1}{2}}\Omega_{\frac{3}{2}.\frac{1}{2}}+3\Omega_{\frac{3}{2},-\frac{3}{2}}\Omega_{\frac{3}{2},\frac{3}{2}}),\\
\label{eq:OO_00}
(\Omega\Omega)_{0,0}&=&\frac{1}{2}(\Omega_{\frac{3}{2},\frac{3}{2}}\Omega_{\frac{3}{2}.-\frac{3}{2}}-\Omega_{\frac{3}{2},\frac{1}{2}}\Omega_{\frac{3}{2}.-\frac{1}{2}}+\Omega_{\frac{3}{2},-\frac{1}{2}}\Omega_{\frac{3}{2}.\frac{1}{2}}-\Omega_{\frac{3}{2},-\frac{3}{2}}\Omega_{\frac{3}{2}.\frac{3}{2}}).
\end{eqnarray}
In this letter, we use Eq.(\ref{eq:OO_00}) to calculate the $S=0$, $S_z=0$ state. 

To extract the  $L=0$ state at the sink $t$ on the lattice,  we employ the cubic-group projection defined by 
\begin{equation}
P_{\nu}^{a}=\frac{d_{a}}{g}\sum_{i}^{g}D_{\nu\nu}^{a}(R_{i})^{*}R_{i},
\end{equation}
where $a$ represents an irreducible representation of the cubic group with dimension $d_{a}$, 
$R_{i}$ is an element of the cubic group acting on $\vec{r}$ of $\Omega$ operators,  $D^{a}(R_{i})$ is the corresponding matrix in the irreducible representation acting on the spin of $\Omega$ operators,
and $g$ is the order of the cubic group. 
We use the $A_{1}$ projection, which generates the $L=0$ as well as the $L=4,6,\cdots$ states, where the $L=4,6,\cdots$ components are expected to be negligibly small.
For example, we have
\begin{equation}
P_\nu^{A_1}  \Omega_{0,0}(\vec r)=  \frac{1}{24}\sum_{i=1}^{24} \Omega_{0,0}(R_{i} \vec r).
\end{equation}

\

\noindent
{\bf \em The $\Omega\Omega$ potential} \ \  
We employ 399 gauge configurations generated by the PACS-CS Collaboration
 with the renormalization group improved gauge action and the non-perturbatively
$\mathcal{O}(a)$ improved Wilson quark action in 2+1 flavor QCD~\cite{aoki:pacscs}.
These configurations were obtained at $\beta=1.90$ ($a=0.0907(13)$ fm) on the $32{}^{3}\times64$
lattice, whose physical extension is $L=2.9$ fm, with the hopping
parameters $\kappa_{ud}=0.13700$ and $\kappa_{s}=0.13640$,
corresponding to $m_{\pi}=701(5)$ MeV and $m_{\Omega}=1966(6)$ MeV.
We employ the wall quark-source with Coulomb gauge fixing. The 
periodic (Dirichlet) boundary condition is imposed in spacial (temporal) direction.
To improve the statistics, we perform the measurement at 64 source time slices for each configuration,
where  the unified contraction algorithm~\cite{doi:2012} is used to calculate the NBS wave functions.
Statistical errors are estimated by the Jack-Knife method.
We make analyses with the bin sizes of 1, 3, 7, 19, 21 and 57,
and the bin size dependence is found to be negligible.
Hereafter, we show the results obtained with the bin size of 1, unless otherwise indicated.

\begin{figure}[tb]
\begin{center}
\includegraphics[angle=270,scale=0.5]{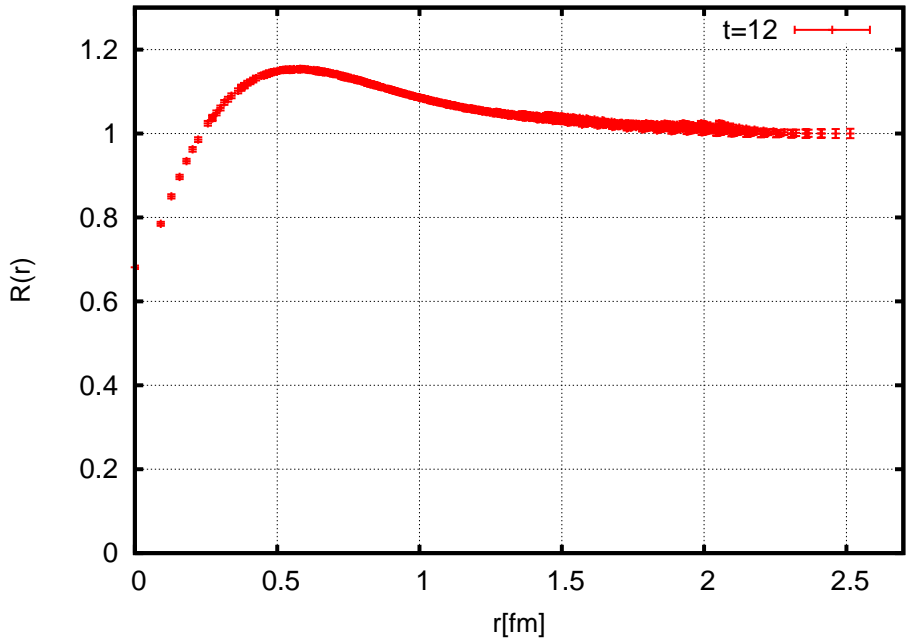}
\caption{The $R$-correlator in the $^1S_0$ channel as a function of the 
 relative distance $r$ for $t-t_{0}=12$.
}
\label{fig:R-corr}
\end{center}
\end{figure}

Fig.~\ref{fig:R-corr}  shows the $R$-correlator (Eq.(\ref{eq:R-col})) 
in the $^1S_0$ channel as a function of $r$ for  $t-t_{0}=12$, where  
the $R$-correlator is normalized to be 1 at $r=2.5$ fm.
 We find  that the $R$-correlator is enhanced at  intermediate distance, while
 it is suppressed at short distance.  The latter behavior is 
 consistent with the partial Pauli blocking in the quark level, similar to the 
 situations in the nucleon-nucleon force \cite{Aoki:2012tk}.

\begin{figure}[tb]
\begin{center}
\includegraphics[angle=270,scale=0.5]{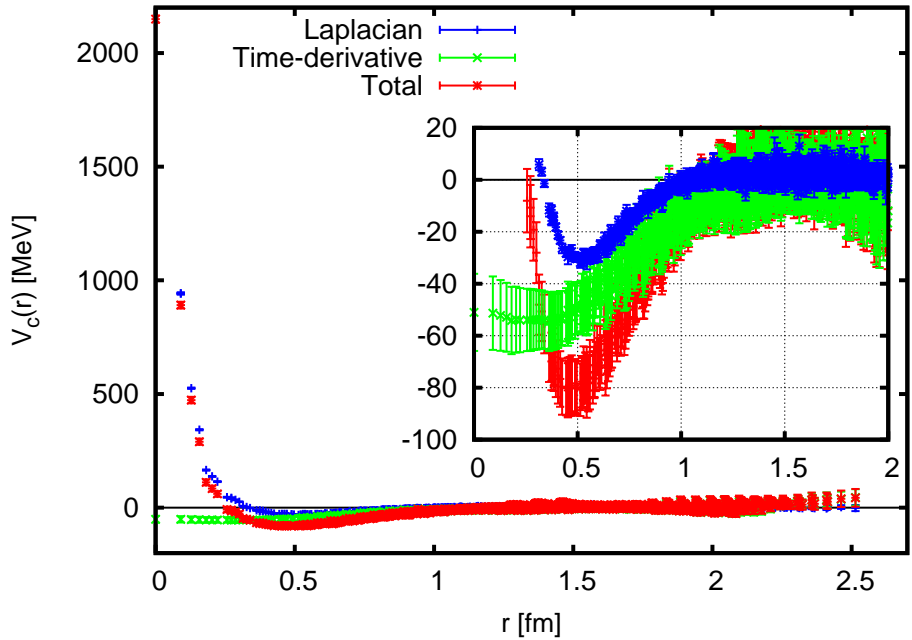}
\caption{The effective central potential for the $\Omega\Omega$ system  in the 
$^1S_0$  channel at $t-t_{0}=12$.  We separately plot the Laplacian term (blue), the time derivative term (green)
and the total (red).}%
\label{FIG:components}
\end{center}
\end{figure}

\begin{figure}[tb]
\begin{center}
\includegraphics[angle=270,scale=0.5]{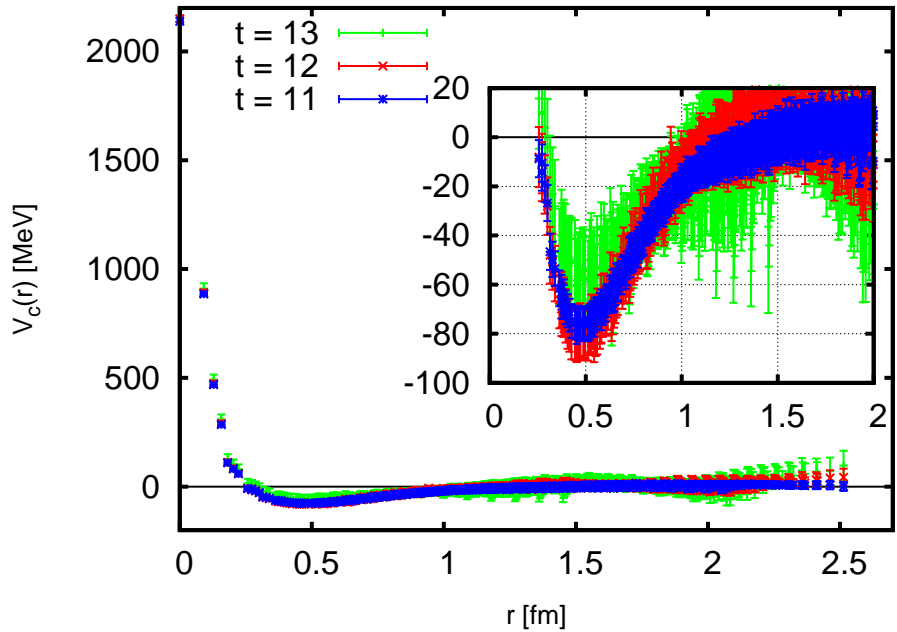}
\caption{The effective central potential $V_{c}(r)$ 
 in the $^{1}S_{0}$ channel at $t-t_{0}=11$ (blue), $12$ (red), $13$ (green). %
}
\label{fig:t-dep}
\end{center}
\end{figure}

Shown in Fig.~\ref{FIG:components} is  the effective central potential $V_{c}(r)$ 
between $\Omega$ baryons  at $t-t_{0}=12$ in the $^1S_0$  channel.  The Laplacian term
and the time derivative term calculated from the $R$-correlator in the right hand side of 
Eq. (\ref{eq:tdep}) are separately plotted in the figure, together with the total potential. 
We here approximate the time derivative term in Eq.(\ref{eq:Potential}) by $\frac{\partial}{\partial t}R(t)=\frac{R(t+1)-R(t-1)}{2}$ and
 $\frac{\partial^{2}}{\partial t^{2}}R(t)=R(t+1)+R(t-1)-2R(t)$.
We find that the  time derivative terms have sizable contributions to the total potential:
 The 1st derivative in $t$ dominates over the 2nd derivative in $t$. The latter 
 corresponding to the relativistic effect is negligible.
 Fig.~\ref{fig:t-dep}  shows the time dependence of $V_{c}(r)$  at $t-t_{0}=11$,
$12$, $13$. This particular region of $t$ is chosen to suppress
 contaminations from excited states in the single $\Omega$ propagator at smaller $t$ and simultaneously to avoid statistical errors at larger $t$.
  We observe that the potential is nearly independent of $t$
 within statistical errors, as expected in 
 the  time-dependent method \cite{Ishii:2012plb}.
 
 The effective central potential $V_{c}(r)$ in Fig.~\ref{fig:t-dep} 
 has a repulsive core at short
 distance and an attractive well at intermediate distance. This  is qualitatively similar to
 the $NN$ and $\Xi\Xi$ interactions, where the partial Pauli blocking in the 
 quark level takes place and the $R$-correlators are suppressed at short distance.
 On the other hand, 
  no Pauli blocking is active for the $H$ dibaryon and the $N\Omega$ dibaryon, so that
   there is no repulsive core in these cases   \cite{Inoue:2011ai,Etminan:2014tya}. 
  
\

\noindent
{\bf \em The $\Omega\Omega$ phase shift} \ \  
To obtain the $S$-wave phase shift, we fit the potential in Fig.\ref{fig:t-dep}
using a function which contains two Gaussian terms plus the Yukawa squared  term  with a form factor \cite{Aoki:2012tk}, given by
\begin{equation}
V_c(r)=a_{1}e^{-a_{2}r^{2}}+a_{3}e^{-a_{4}r^{2}}+a_{5}(1-e^{-a_{6}r^{2}})^{2}(\frac{e^{-a_{7}r}}{r})^{2},\qquad\lim_{r\rightarrow0}V(r)=a_{1}+a_{3}.
\end{equation}
This ${\rm 2\ Gauss}+({\rm Yukawa})^{2}$ form gives a fairly good fit with  $\chi^{2}/{\rm d.o.f}\sim 0.50$
at $t-t_{0}=12$. 
The resulting parameters are 
$a_{1}={\rm 1.69(6)\times10^{3}\, MeV}$, $a_{2}=1.24(3)\times10^{2}\, {\rm fm}^{-2}$, $a_{3}=4.44(68)\times10^{2}\,{\rm MeV}$, $a_{4}=5.68(1.31)\, {\rm fm}^{-2}$, $a_{5}=-7.06(14.64)\times10^{4}\, {\rm MeV}$, $a_{6}=6.25(5.77)\times10^{-1}\, {\rm fm}^{-2}$, $a_{7}=3.43(30)\, {\rm fm}^{-1}$ at $t-t_{0}=12$.

Using the fitted potential, we solve the Schr\"{o}dinger
equation in the infinite volume to calculate the scattering phase
shift $\delta_L(k)$ of the $\Omega\Omega$ system in $^1S_0$ channel by  the formula with $L=0$,
\begin{equation}
\tan\delta_{L}(k)=\lim_{x_{1},x_{2}\rightarrow\infty}\frac{\psi_{k}(x_{2})\sin(kx_{1}-\frac{L\pi}{2})-\psi_{k}(x_{1})
\sin(kx_{2}-\frac{L\pi}{2})}{\psi_{k}(x_{1})\cos(kx_{2}-\frac{L\pi}{2})-\psi_{k}(x_{2})\cos(kx_{1}-\frac{L\pi}{2})},
\end{equation}
where $\psi_{k}$ is the wave function and $k$ is the magnitude of 
the momentum.

\begin{figure}[tb]
\begin{center}
\includegraphics[angle=270,scale=0.5]{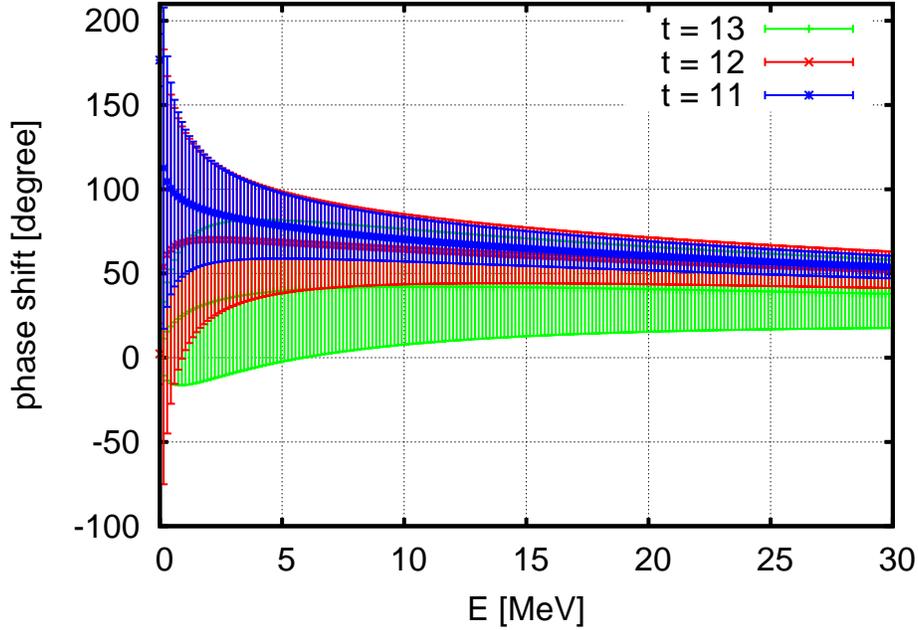}
\caption{Phase shift $\delta_0(k)$ of the $\Omega\Omega$ in the ($^{1}S_{0}$) channel 
at $t-t_{0}=11$ (blue), $12$ (red), $13$ (green).%
}
\label{fig:phase}
\end{center}
\end{figure}

Fig.~\ref{fig:phase} shows the phase shift as a function of the kinetic energy, $E=k^2/m_\Omega$.
The result
  indicates that the $\Omega\Omega$ interaction is attractive at low energies, while the existence of the bound state is inconclusive because of large  statistical errors near $k=0$. Indeed, the 
 central value of $\delta_0(k=0)$ at $t-t_0=11$ is  $180^{\circ}$, 
 while it becomes zero  at $t-t_0=12, 13$. 
  Due to the same reason,
 the scattering length and the effective range cannot be extracted reliably from this phase shift.
 A possible physical interpretation of such a situation is that the  
$\Omega\Omega$ system at the present quark masses 
  may appear close to the  unitary limit where the scattering length diverges and changes its sign.

\
 
\noindent
{\bf \em  Conclusion} \ \ 
In this letter, we have calculated the effective central potential and the  scattering
phase shift between $\Omega$ baryons 
 in the  $^1S_0$
channel by using the HAL QCD method. 
We have found that the $\Omega\Omega$ potential has short range repulsion and 
intermediate range attraction just like the nucleon-nucleon potential. 
The short range repulsion of this system is a reflection of the partial  Pauli blocking in the 
quark level similar to the nucleon-nucleon potential, 
and is in contrast to the cases of the  $H$-dibaryon or $N\Omega$ system with no repulsion.
 The $\Omega\Omega$ interaction is attractive  at low energies, but
  is not strong enough to form a tightly bound dibaryon at 
quark masses corresponding to 
$m_\pi\simeq 700$ MeV and $m_\Omega \simeq 1970$ MeV.
   Rather, the system may  appear close to the  unitary limit at these quark masses.
 We plan to carry out the 2+1 flavor simulations
at the physical quark masses, in order to investigate whether the attraction found in the
present study increases or decreases toward the physical quark masses.

\

\noindent
{\em Acknowledgements} \ \ 
We are grateful for authors and maintainers of CPS++ \cite{CPS}, a modified
version of which is used for simulations done in this work. 
We thank PACS-CS Collaboration and ILDG/JLDG
for providing us the 2+1 flavor gauge configurations \cite{JLDG}.
Numerical computations of this work have been carried out by the KEK
supercomputer system (BG/Q) under JICFuS-H26-3 and by local machines at University of Tsukuba.
This work is supported in part by 
the Grant-in-Aid of the Ministry of Education, Science and Technology, Sports and Culture 
(Nos. 20340047, 22540268, 19540261, (B)25287046, (C)26400281, 24740146) and the  Strategic program for Innovative Research (SPIRE) Field 5,
and  JICFuS \cite{JICFUS}.
T.H. was partially supported by RIKEN iTHES Research Group.

%%%%%%%%%%%%% ref %%%%%%%%%%%%%%%

\end{document}